\documentclass[12pt]{article}
\usepackage{amsfonts}
\usepackage{amsmath}
\usepackage{geometry}
\usepackage{amssymb}
\usepackage{lmodern}

\newtheorem{theorem}{Theorem}

\newtheorem{assumption}{Assumption}

\newtheorem{corollary}{Corollary}

\newtheorem{example}{Example}

\newtheorem{proposition}{Proposition}
\newtheorem{remark}{Remark}

\begin{document}

\title{{\LARGE Unobserved Consideration and Consumer Welfare}}
\author{{\large Debopam Bhattacharya\thanks{%
Bhattacharya is grateful to Tatiana Komarova for discussions. Financial
support from ESRC grant number ES/W010461/1 is gratefully acknowledged.}} \\
{\large University of Cambridge} \and {\large Ryan J. Martin\thanks{%
Martin is grateful to his UCLA advisors, Rosa Matzkin, John Asker, Denis
Chetverikov, Jinyong Hahn, and Elisabeth Honka, and to faculty and seminar
participants at UCLA. He is also grateful to former colleagues at the Bank
of Canada for helpful discussions, especially Kim Huynh and Oleksandr
Shcherbakov. This paper subsumes some results from his doctoral
dissertation, titled ``Consumer Welfare in Online Markets'' (UCLA, 2019).
The views expressed here are his own and do not necessarily reflect those of
Faire Wholesale, Inc. or the Bank of Canada.}} \\
{\large Faire Wholesale, Inc.}}
\date{{\large 4 November, 2026}}
\maketitle

\textbf{Abstract}: Can welfare effects of price change be identified when
consumers' consideration sets are heterogeneous and unobserved? We establish
that the welfare distribution is the same functional of structural choice
probabilities under stable but limited consideration as under full
consideration. The result permits unrestricted income-effects, preference
heterogeneity, and dependence between preferences and consideration. Within
this class, limited consideration admits a full-consideration representation
preserving choice-probabilities and money-metric compensation:
consideration/preferences can remain unidentified while demand and welfare
are identified. When consideration sets contract as prices rise, the welfare
distribution can be bounded using demand and a bound on choice-set attrition.

\section{Introduction}

When evaluating the welfare effects of a tax or subsidy, must an economist
know which products consumers actually considered? Indeed, consumers may
fail to consider all available products because of limited awareness, search
costs, or cognitive constraints, cf. Honka et al 2019. Consequently, an
unchosen product may be unattractive or simply overlooked. A central
objective of the consideration-set literature in economics and marketing is
to distinguish these explanations through restrictions on preferences,
search, or attention, or through variation in choice menus. Such
restrictions, however, limit the settings in which this distinction is
achievable, potentially complicating welfare analysis. For example, Crawford
et al. (2021) argue that even when preference parameters are identified,
unknown choice sets can obstruct the evaluation of counterfactual choice
probabilities, price elasticities, and consumer surplus. Indeed, unobserved
consideration would appear to obstruct welfare identification because the
welfare loss from price increase of an option would differ, depending on
whether or not it belonged to the consumer's consideration set and on which
substitutes she would consider, something observed purchases do not directly
reveal.

Bhattacharya (2015, 2018) established that, under full consideration, the
distribution of compensating variation for simultaneous price changes is
identified from demand, i.e. structural choice probabilities, allowing
unrestricted income effects and preference heterogeneity of arbitrary
dimension and distribution. The question we address here is whether
unobserved consideration introduces an additional identification challenge
for welfare analysis. Our main result shows that, when consideration sets
remain stable, the same functional of demand identifies the entire
distribution of compensating variation without requiring identification of
individual consideration sets or their distribution, even when consideration
and preferences are arbitrarily correlated. Consideration affects welfare
through its effect on structural demand, which can, given suitable
identifying variation, be estimated \textit{without} separately recovering
preferences and consideration sets. Thus, once structural demand is
identified, no additional information about consideration is needed to
recover welfare. This conclusion is not immediate: unlike preference
heterogeneity, limited consideration restricts the alternatives over which
consumers maximize utility, while the welfare effects of price changes
depend on their ability to substitute towards alternatives within their
consideration sets.

We also establish a constructive equivalence result that helps explain the
welfare conclusion. Under stable consideration, every limited-consideration
model in our preference class admits a full-consideration representation
that preserves structural demand. Thus, although preferences and
consideration sets may remain observationally indistinguishable, the welfare
distribution can nevertheless be identified.

For example, consider automobile demand in the spirit of Berry et al.
(1995). Recovering consumers' willingness to pay for fuel efficiency
requires distinguishing preferences from consideration: low sales of a new
fuel-efficient model may reflect a low valuation of fuel efficiency or its
exclusion from consumers' consideration sets, perhaps because they were
unaware of the model. Our result shows that this distinction need not be
resolved to evaluate the welfare effects of a tax or subsidy that changes
vehicle prices. Provided consideration remains stable and structural demand
is identified at the required price--income configurations, the entire
distribution of compensating variation is identified without separately
recovering valuations of vehicle characteristics or consideration sets.

The equivalence is constructive. Every stable limited-consideration model
within the maintained preference class admits a full-consideration
representation in which excluded alternatives receive finite, strictly
income-increasing utilities whose ranges lie below those of considered
alternatives. This construction preserves choices and individual
compensating variation across the relevant price and income configurations.
The result therefore establishes more than observational equivalence: it
shows that ambiguity about whether choice behavior reflects preferences or
consideration need not translate into ambiguity about welfare.

Three policy examples illustrate the relevance of this distinction. The
United Kingdom imposed a 20\% value-added tax (VAT) on private-school fees
from 1 January 2025 (HM Revenue \& Customs, 2024); Edinburgh introduced a
visitor levy for eligible stays from 24 July 2026 (City of Edinburgh
Council, 2026); and US clean-vehicle purchase credits ended for vehicles
acquired after 30 September 2025 (Internal Revenue Service, 2026; Allcott et
al., 2026; Buckberg and Cole, 2025). Each intervention, by design, changed
the consumer prices of multiple alternatives. Yet, data on enrolment,
accommodation bookings, or vehicle sales typically do not reveal which
schools, properties, or car models each consumer considered. Our result
implies that such information is unnecessary for welfare calculations if
consideration remains stable.

Our equivalence result has a logical boundary. If a sufficiently large
subsidy makes an alternative preferable to every competitor for almost every
consumer, its limiting demand identifies its consideration probability.
Conditional on rationalizability (cf. Bhattacharya, 2021), under this
additional restriction, a limit below one rules out full consideration. The
welfare formula, however, remains valid with or without this restriction.

Finally, we examine a setting in which consideration is not stable but
contracts as prices rise. This may arise, for example, when online consumers
screen out products exceeding price thresholds. In this case, the demand
expression that identifies the welfare distribution under stable
consideration becomes an upper bound on its cumulative distribution
function, while an upper bound on the fraction of consumers whose
consideration sets change yields a lower bound. Thus, although endogenous
consideration can prevent point identification, demand and limited
information about consideration-set attrition still provide welfare bounds.

The rest of the paper is organized as follows. Section 2 sets out the
environment and assumptions. Section 3 relates the analysis to the existing
literature. Section 4 derives the distribution of ex-post compensation for
simultaneous price changes and develops the policy examples. Section 5
establishes observational and welfare equivalence between stable limited and
full consideration and characterizes its boundary under eventual dominance.
Section 6 derives welfare bounds under contracting consideration. The
appendices distinguish ex-ante from ex-post compensation and extend the
analysis to additions and removals of alternatives.

\section{Setting and assumptions}

\label{sec:setting}

There are $J\geq 2$ alternatives indexed by $\mathcal{J}=\{0,1,\ldots ,J-1\}$%
. Write $p=(p_{0},\ldots ,p_{J-1})$ and $p^{\prime }=(p_{0}^{\prime },\ldots
,p_{J-1}^{\prime })$ for initial and counterfactual prices. Income is $y$.
Define 
\begin{equation}
\Delta _{j}=p_{j}^{\prime }-p_{j},\qquad j\in \mathcal{J}.
\label{eq:increments}
\end{equation}%
Without loss of generality, relabel alternatives so that 
\begin{equation}
\Delta _{0}\leq \Delta _{1}\leq \cdots \leq \Delta _{J-1}.
\label{eq:ordered}
\end{equation}

A consumer is described by tastes $\eta $ and a nonempty consideration set $%
S\subseteq \mathcal{J}$. Probabilities below refer to a fixed population
distribution of $(\eta ,S)$, with any maintained observable conditioning
variables suppressed for notational economy.

\begin{assumption}[Preferences and domain]
\label{ass:preferences} For each $j\in \mathcal{J}$, utility is $%
U_{j}(m;\eta )$, where $m$ denotes residual income $y-p_{j}$ where $y$ is
consumer budget. For every individual taste vector $\eta $, $U_{j}(\cdot
;\eta )$ is finite, continuous, and strictly increasing on $\mathbb{R}$.
\end{assumption}

Note that no quasilinearity, common marginal utility of income, additive
heterogeneity, or independence across alternative-specific tastes is imposed.

\begin{assumption}[Stable consideration and population]
\label{ass:stable} Each consumer's realized set $S$ and the joint population
distribution of $(\eta ,S)$ are unchanged between the baseline and all
constructed price--income configurations used below.
\end{assumption}

Assumption \ref{ass:stable} permits arbitrary dependence between $S$ and $%
\eta $. It excludes consideration responses to the price and income changes
being evaluated. Stable consideration is plausible when consumers choose
from established shortlists that are costly to revise relative to the price
change (Hauser and Wernerfelt, 1990). Empirical evidence from US auto
insurance and banking suggests that consumers select the providers to
investigate before observing their quotes or offered interest rates (Honka
and Chintagunta, 2017; Honka et al., 2017). Experimental evidence also shows
that information differentiating brands can increase the stability of
consideration sets across purchase occasions (Mitra, 1995). For example,
travellers typically search for hotel accommodation through one or two fixed
platforms (Expedia, Booking.com etc.), with not every property listed on
every platform. A price increase for a listed property is unlikely to make
them switch to a different platform. Similarly, a family may shortlist a few
private schools after visiting campuses, assessing suitability, and
establishing feasible travel arrangements. A differential fee increase could
change its preferred school within that shortlist without making the cost of
investigating additional schools outside the shortlist worthwhile. Even when
price changes are large enough to alter consideration sets, calculating
compensation holding initial consideration sets fixed remains a useful
benchmark. It isolates the welfare consequences of changed prices within
consumers' initial opportunity sets, abstracting from further adjustments in
search or attention. In appendix B below, we allow effective choice sets to
change through the addition or removal of available alternatives, while
holding latent consideration over the union menu fixed.

For a price vector $p=(p_{0},\ldots ,p_{J-1})$, define indirect utility and
choice by 
\begin{equation}
\begin{array}{rl}
V_{S}(p,y;\eta ) & =\max_{j\in S}U_{j}(y-p_{j};\eta ), \\ 
d(p,y;\eta ,S) & \in \mathop{\rm arg\,max}_{j\in S}U_{j}(y-p_{j};\eta ).%
\end{array}
\label{eq:choice}
\end{equation}%
Under the no-ties assumption below, the maximizing alternative is unique
with probability one at each fixed configuration used in the analysis. A
fixed measurable rule selects an alternative on any null event of ties.
Population choice probabilities are 
\begin{equation}
q_{j}(p,y)=\Pr \{d(p,y;\eta ,S)=j\},\qquad j\in \mathcal{J}.
\label{eq:probabilities}
\end{equation}%
Thus $\sum_{j\in \mathcal{J}}q_{j}(p,y)=1$.

\begin{assumption}[No ties]
\label{ass:noties} At every fixed price--income configuration $(p,y)$ used
in the analysis, the utility-maximizing alternative in the consumer's
consideration set is unique with probability one. Formally, 
\begin{equation*}
\Pr \left\{ \left\vert \mathop{\rm arg\,max}_{j\in S}U_{j}(y-p_{j};\eta
)\right\vert =1\right\} =1.
\end{equation*}%
The probability is taken over the joint distribution of $(\eta ,S)$.
\end{assumption}

This is the standard no-ties regularity condition: ties for the highest
utility occur with probability zero at any fixed configuration, although an
individual may be indifferent at an individual-specific configuration. The
condition is pointwise in $(p,y)$; it does not require a single
probability-one set of individuals with unique choices at every
configuration simultaneously.

\section{Relation to existing literature}

\label{sec:literature}

A substantial literature in Marketing studies consideration sets in
empirical settings. See, for example, Honka et al. (2019) for a
comprehensive review of this literature. In econometrics, Manski (1977)
included unobserved choice sets within a general random utility framework.
Subsequent identification results have typically restricted preferences,
consideration or the available variation in order to separate preferences
from consideration. For example, Barseghyan et al. (2021a) combine
preference restrictions with a lower bound on choice-set size; Barseghyan et
al. (2021b) exploit a specified scalar-index preference structure satisfying
single crossing, together with suitable covariate variation and
consideration assumptions. The utilities specified below may or may not
satisfy these restrictions. The identification results of the abovementioned
studies are therefore compatible with observational equivalence in the
broader class studied here. Corollary \ref{cor:reject-full} illustrates
explicitly how an additional preference restriction, combined with
sufficiently rich price support, can separate the two model classes.

Abaluck and Adams-Prassl (2021) exploit symmetry restrictions on
cross-attribute demand responses under full consideration and departures
from these restrictions when attributes affect consideration. These
restrictions depend on their maintained utility structure including additive
separability of a utility characteristic across alternatives. However, with
unrestricted income effects, uncompensated cross-price symmetry -- the key
to their approach -- need not hold under full consideration. The following
example illustrates the point.

\textbf{Example}: Consider three alternatives $1,2,3$; take alternative 3 as
the outside option with $p_{3}=0$, and suppose every consumer considers all
three alternatives. For a parameter $\lambda \geq 0$, let utilities be given
by%
\begin{equation}
\begin{array}{rl}
U_{1}(m;\varepsilon _{1}) & =(1+\lambda )m+\varepsilon _{1}, \\ 
U_{2}(m;\varepsilon _{2}) & =m+\varepsilon _{2}, \\ 
U_{3}(m;\varepsilon _{3}) & =m+\varepsilon _{3}.%
\end{array}
\label{eq:income-symmetry-utilities}
\end{equation}%
The shocks are independent standard type-I extreme-value random variables,
independent of prices and income. All utilities are continuous and strictly
increasing on $\mathbb{R}$, and ties have probability zero. Subtracting the
common term $y$ from utilities gives deterministic indices 
\begin{equation*}
v_{1}=\lambda y-(1+\lambda )p_{1},\qquad v_{2}=-p_{2},\qquad v_{3}=0.
\end{equation*}%
Consequently, 
\begin{equation}
q_{j}(p,y;\lambda )=\frac{\exp (v_{j})}{\sum_{k=1}^{3}\exp (v_{k})},\qquad
j=1,2,3.  \label{eq:income-symmetry-shares}
\end{equation}%
Holding income fixed, the cross-price responses for the two inside goods are 
\begin{equation}
\frac{\partial q_{1}}{\partial p_{2}}=q_{1}q_{2},\qquad \frac{\partial q_{2}%
}{\partial p_{1}}=(1+\lambda )q_{1}q_{2}.  \label{eq:income-symmetry-gap}
\end{equation}%
For every $\lambda >0$ and finite prices and income, the latter cross-price
derivative minus the former is $\lambda q_{1}q_{2}>0$, despite full
consideration.

Now eliminate income effects by imposing $\lambda=0$. Within this family,
this restriction is necessary and sufficient for all choice probabilities to
be independent of income. Utilities become $U_j(m;\varepsilon_j)
=m+\varepsilon_j$ for every alternative, so that 
\begin{equation*}
q_1=\frac{e^{-p_1}}{1+e^{-p_1}+e^{-p_2}},\qquad q_2=\frac{e^{-p_2}}{%
1+e^{-p_1}+e^{-p_2}},\qquad q_3=\frac{1}{1+e^{-p_1}+e^{-p_2}}.
\end{equation*}
The symmetry restriction is then restored: 
\begin{equation}
\frac{\partial q_1}{\partial p_2} =\frac{\partial q_2}{\partial p_1}=q_1q_2.
\label{eq:income-symmetry-restored}
\end{equation}
Thus the same family exhibits asymmetric uncompensated price responses with
income effects and symmetric responses when income effects are ruled out by
assumption.

The above conclusion is not peculiar to logit errors. Under the sufficient
no-income-effects restriction $U_{j}(m;\eta )=m+a_{j}(\eta )$, let 
\begin{equation*}
H(p)=\mathbb{E}\left[ \max_{j\in S}\{a_{j}(\eta )-p_{j}\}\right] .
\end{equation*}%
If this expectation is finite, differentiation under the expectation is
valid, and $H$ is twice continuously differentiable, then $q_{j}=-\partial
H/\partial p_{j}$ and equality of mixed partial derivatives implies
symmetric cross-price responses. This argument applies both to full
consideration and to stable limited consideration, with arbitrary dependence
between $S$ and $\eta $. Hence symmetry alone does not establish full
consideration, even in this no-income-effects benchmark. Conversely, the
example with $\lambda >0$ shows that asymmetry can arise under full
consideration when income effects are allowed. Our framework, in contrast,
allows arbitrary income effects.

Cattaneo et al. (2020) consider partial identification of a fixed preference
ordering from choice probabilities across varying menus. In contrast,
heterogeneous preferences are central to our exercise and the goal is to
identify from a single cross-section the distribution of welfare from
structural demand without identifying preferences or consideration.

Crawford et al. (2021) survey methods for preference estimation with
unobserved choice-set heterogeneity. Their discussion highlights that
identifying preference parameters need not suffice to identify choice
probabilities or consumer surplus when the distribution of choice sets
remains unknown. Our analysis addresses a complementary question: given
structural choice probabilities, is separate identification of consideration
sets necessary to recover the distribution of welfare effects? In terms of
our notation above, Crawford et al. (2021) point out that when utilities
take a parametric form $U_{k}(m;\eta )=V_{k}(m;\theta ,\eta )$, knowing
preference parameters $\theta $ alone makes it impossible to identify%
\begin{equation}
q_{j}(p,y)=\Pr \{j=\arg \max_{k\in S}V_{k}(y-p_{k};\theta ,\eta )\},
\end{equation}%
because these probabilities also depend on the joint distribution of $\left(
S,\eta \right) $. Our result, instead, says that knowing the functions $%
q_{j}(\cdot ,\cdot )$ can identify a welfare functional without identifying
the joint distribution of $\left( S,\eta \right) $.

Our paper is also related to Allen 2026, who studies limited consideration
within a restrictive \textit{additive} random utility framework. His
identification result pertains to differences in indirect utility, measured
in the units of the specified utility indices. Such differences are \textit{%
not} invariant to general monotonic transformations of utility and do not,
without additional assumptions, have a monetary interpretation. This
distinction is especially important when consumers have heterogeneous
marginal utilities of income, because utility changes cannot generally be
translated into monetary welfare using a common conversion factor. Further,
Allen's proof relies on the additive preference structure, which allows
choice probabilities to be expressed as derivatives of average indirect
utility. In contrast, our analysis (a) allows preference heterogeneity to
have any arbitrary dimension and enter utility in any arbitrary way, (b)
pertains to money-metric compensating variation, which is invariant to
monotonic transformations of utility, (c) identifies its entire distribution
from structural demand while allowing unrestricted income effects and
unobserved consideration sets, and (d) studies how welfare identification
changes when consideration sets respond to prices.

\section{Welfare from structural demand}

\label{sec:welfare}

We study ex-post compensating variation: the transfer that restores a
consumer's baseline utility after prices change, holding the realized
consideration set fixed. For the change from $p$ to $p^{\prime }$, define $%
C=C(S,\eta )$ by 
\begin{equation}
V_{S}(p^{\prime },y+C;\eta )=V_{S}(p,y;\eta ).  \label{eq:cv-definition}
\end{equation}%
Its distribution is taken over the joint distribution of preferences and
consideration sets. Appendix~\ref{app:compensation} distinguishes this
measure from ex-ante compensation, defined before the consideration set is
realized, and explains the additional cardinal utility structure required by
the latter.

\subsection{The welfare distribution}

\begin{proposition}[Existence, uniqueness, and support]
\label{prop:support} Under Assumptions \ref{ass:preferences}--\ref%
{ass:stable}, the ex-post compensating variation exists uniquely and
satisfies 
\begin{equation}
\Delta _{0}\leq \min_{j\in S}\Delta _{j}\leq C\leq \max_{j\in S}\Delta
_{j}\leq \Delta _{J-1}  \label{eq:support}
\end{equation}%
for every individual. For every $a\in \mathbb{R}$, 
\begin{equation}
\{C\leq a\}=\{V_{S}(p^{\prime },y+a;\eta )\geq V_{S}(p,y;\eta )\}.
\label{eq:event}
\end{equation}
\end{proposition}

\noindent\textbf{Proof.} For fixed $(\eta,S)$, define $g(c)=\max_{j\in
S}U_j(y+c-p^{\prime }_j;\eta)$. A finite maximum of continuous functions is
continuous. It is also strictly increasing: if $c_2>c_1$ and $j_1$ maximizes
at $c_1$, then 
\begin{equation*}
g(c_2)\geq U_{j_1}(y+c_2-p^{\prime }_{j_1};\eta) >U_{j_1}(y+c_1-p^{\prime
}_{j_1};\eta)=g(c_1).
\end{equation*}
Let $\ell_S=\min_{j\in S}\Delta_j$ and $h_S=\max_{j\in S}\Delta_j$. For
every $j\in S$, 
\begin{equation*}
y+\ell_S-p^{\prime }_j\leq y-p_j\leq y+h_S-p^{\prime }_j.
\end{equation*}
Monotonicity and maximization imply 
\begin{equation*}
g(\ell_S)\leq V_S(p,y;\eta)\leq g(h_S).
\end{equation*}
Continuity and strict monotonicity establish existence, uniqueness, and (\ref%
{eq:support}), including when $\ell_S=h_S$. They also establish (\ref%
{eq:event}). The population bounds follow because $S\subseteq\mathcal{J}$.
\hfill$\square$

At a candidate transfer $a$, alternative $j$ has residual income $y-p_{j}$
at baseline and $y+a-p_{j}^{\prime }$ after the intervention. Construct a
budget with residual income 
\begin{equation*}
\max \{y-p_{j},y+a-p_{j}^{\prime }\}
\end{equation*}%
for each alternative. If the chosen alternative at this budget is one whose
compensated residual income weakly exceeds its baseline value, compensation
is sufficient. If its baseline residual income is strictly larger,
compensation is insufficient. Strict monotonicity and the no-ties condition
make this classification exact. Crucially, it uses the same realized set
throughout and does not require that set to be observed.

For $k\in \{0,\ldots ,J-2\}$ and $\Delta _{k}\leq a<\Delta _{k+1}$, define
the constructed prices and income by 
\begin{eqnarray}
r_{j}^{(k)}(a)& = &\left\{%
\begin{array}{ll}
p_{j}^{\prime }-\min \{a,0\}, & j\leq k, \\ 
p_{j}+a-\min \{a,0\}, & j>k,%
\end{array}%
\right.\qquad j\in \mathcal{J},  \label{eq:r} \\
y(a)& = &y+a-\min \{a,0\}.  \label{eq:w}
\end{eqnarray}%
The outside-option price is zero: if $a\geq 0$, the outside option lies
among $0,\ldots ,k$; if $a<0$, it lies among $k+1,\ldots ,J-1$. For
nonnegative $a$, the construction simplifies to 
\begin{equation*}
r^{(k)}(a)=(p_{0}^{\prime },\ldots ,p_{k}^{\prime },p_{k+1}+a,\ldots
,p_{J-1}+a),\qquad y(a)=y+a.
\end{equation*}

\begin{theorem}[Distribution of ex-post compensating variation]
\label{thm:cdf} Under Assumptions \ref{ass:preferences}--\ref{ass:noties},
and the labeling convention (\ref{eq:ordered}), 
\begin{equation}
F_{C}(a;p,p^{\prime },y)=\left\{ 
\begin{array}{ll}
0, & a<\Delta _{0}, \\[4pt] 
\displaystyle\sum_{j=0}^{k}q_{j}(r^{(k)}(a),y(a)), & \Delta _{k}\leq
a<\Delta _{k+1},\quad k=0,\ldots ,J-2, \\[6pt] 
1, & a\geq \Delta _{J-1}.%
\end{array}%
\right.  \label{eq:cdf}
\end{equation}
\end{theorem}

\noindent \textbf{Proof.} The two outer branches $a<\Delta _{0}$ and $a\geq
\Delta _{J-1}$ follow from Proposition \ref{prop:support}. Next, fix $k$ and 
$a\in \lbrack \Delta _{k},\Delta _{k+1})$ in a nonempty interval. By
Assumption \ref{ass:noties}, the maximizing alternative at $%
(r^{(k)}(a),y(a)) $ is unique almost surely. For an individual $(\eta ,S)$
outside this null set, write, for every $j\in S$, 
\begin{equation*}
b_{j}=U_{j}(y-p_{j};\eta ),\qquad n_{j}=U_{j}(y+a-p_{j}^{\prime };\eta ).
\end{equation*}%
Ordering and strict monotonicity imply 
\begin{equation*}
n_{j}\geq b_{j}\quad \hbox{if }j\leq k,\qquad n_{j}<b_{j}\quad \hbox{if }j>k.
\end{equation*}%
By (\ref{eq:r})--(\ref{eq:w}), the constructed utility is 
\begin{equation*}
h_{j}:=U_{j}(y(a)-r_{j}^{(k)}(a);\eta )=\left\{ 
\begin{array}{ll}
n_{j}, & j\leq k, \\ 
b_{j}, & j>k.%
\end{array}%
\right.
\end{equation*}%
Thus $h_{j}=\max \{b_{j},n_{j}\}$. Let $j^{\ast }$ be its unique maximizer
over $S$. If $j^{\ast }\leq k$, then 
\begin{equation*}
\max_{j\in S}n_{j}\geq n_{j^{\ast }}=h_{j^{\ast }}=\max_{j\in S}h_{j}\geq
\max_{j\in S}b_{j}.
\end{equation*}%
If $j^{\ast }>k$, then $h_{j^{\ast }}=b_{j^{\ast }}>n_{j^{\ast }}$ and, for
every $j\neq j^{\ast }$ in $S$, uniqueness implies $n_{j}\leq
h_{j}<h_{j^{\ast }}$. Finiteness of $S$ therefore gives 
\begin{equation*}
\max_{j\in S}n_{j}<h_{j^{\ast }}=\max_{j\in S}b_{j}.
\end{equation*}%
Using (\ref{eq:event}), it follows that, almost surely for this fixed $a$, 
\begin{equation*}
\{C\leq a\}=\{d(r^{(k)}(a),y(a);\eta ,S)\in \{0,\ldots ,k\}\}.
\end{equation*}%
Taking probabilities proves the middle branch of (\ref{eq:cdf}), i.e. for
the case $\Delta _{k}\leq a<\Delta _{k+1}$. Since $a$ was arbitrary, the
identity holds throughout every nonempty interval. \hfill $\square $

The argument of the proof permits unrestricted income effects, completely
general preference heterogeneity and arbitrary dependence between
preferences and consideration sets. The choice probabilities in (\ref{eq:cdf}%
) already incorporate limited consideration, so separate identification of
the consideration-set probabilities is unnecessary.

Equation (\ref{eq:cdf}) has the same functional form as in the
full-consideration result in Bhattacharya (2018). Under limited but stable
consideration, the formula is unchanged, but the choice probabilities
incorporate consideration. This establishes that the ambiguity between
exclusion and unattractive preferences does not translate into ambiguity
about this welfare distribution. Proposition~\ref{prop:equivalence} explains
why, by constructing a full-consideration representation that preserves
individual choices and compensation.

Appendix~\ref{sec:menus} extends the argument to specified additions and
removals of alternatives. That extension uses structural demand under the
union of the initial and final menus, with potential consideration held
fixed across those menus.

\begin{remark}
Note that the absence of ties between distinct alternatives does not imply a
continuous welfare distribution. If $S=\{j\}$, strict monotonicity gives $%
C=\Delta _{j}$ exactly. Thus 
\begin{equation*}
\Pr (C=\Delta _{j})\geq \Pr (S=\{j\}).
\end{equation*}%
Consumers with larger consideration sets may also contribute to an atom.
When all inside-option increments are strictly positive, $%
F_{C}(0)=q_{o}(p,y) $: compensation is zero precisely for consumers choosing
the outside option initially, up to the maintained no-ties condition. When
some inside-option increments are zero, their choice probabilities must be
included as well. The half-open intervals in (\ref{eq:cdf}) handle these
endpoints.
\end{remark}

For each threshold $a$, equation (\ref{eq:cdf}) identifies the CDF of the
required compensation at $a$, provided the structural choice probabilities
at $(r^{(k)}(a),y(a))$ are identified, without the need to identify
individual tastes or consideration sets. The formula for CDFs implies those
for welfare quantiles and, from the bounded support, mean compensation:%
\begin{equation*}
\mathbb{E}C=\Delta _{0}+\int_{\Delta _{0}}^{\Delta _{J-1}}[1-F_{C}(a)]\,da.
\end{equation*}

\subsection{Identifying the demand used in the formula}

\label{subsec:demand-identification} The above theorem is an identification
result conditional on structural demand along the paths $(r^{(k)}(a),y(a))$
entering (\ref{eq:cdf}). Let $X$ collect observed characteristics, $P$
observed prices, $Y$ income, and $D$ the chosen alternative. Suppose $(P,Y)$
is independent of $(\eta ,S)$ conditional on a \textquotedblleft control
function\textquotedblright\ (which can be $X$ or the residual from a
first-stage regression on instruments (cf. Imbens and Newey, 2009)), and
consideration is stable over the relevant configurations. Then the
structural choice probabilities can be identified on the conditional
support. Importantly, limited consideration does not introduce additional
identification requirements for the structural choice probabilities entering
the welfare formula; as under full consideration, these probabilities must
be identified at the relevant counterfactual price and income
configurations, but neither preferences nor consideration sets need to be
separately identified.

\section{Empirical content and observational equivalence}

\label{sec:equivalence}

The above welfare formula does not require separate identification of
consideration. We now show that such identification is generally unavailable
within the maintained preference class, even with the entire structural
demand system. This provides a second interpretation of the welfare result:
both structural demand and compensating variation can be reproduced by a
full-consideration model with different preferences. Letting $\mathcal{J}$
denote the full set of alternatives, our construction assigns each original
preference--consideration triplet $\left\{ U,\eta ,S\right\} $ a constructed
preference type $\widetilde{\eta }$, with utility functions $\widetilde{U}$
such that, under \textit{full} consideration, $\left\{ \widetilde{U},%
\widetilde{\eta },\mathcal{J}\right\} $ reproduces the same structural
demand, and hence, individual compensating variation generated by $\left\{
U,S,\eta \right\} $ under limited consideration.

\begin{proposition}[Observational and ex-post welfare equivalence]
\label{prop:equivalence} For any model $\left\{ U,S,\eta \right\} $
satisfying Assumptions \ref{ass:preferences}--\ref{ass:noties}, there exists
a full-consideration model $\left\{ \widetilde{U},\mathcal{J},\widetilde{%
\eta }\right\} $satisfying the same preference and no-ties conditions, with
a fixed distribution of augmented tastes $\widetilde{\eta }$, that generates
the same choices and structural choice probabilities at every configuration
over which consideration and the population are held fixed. The two models
also generate the same individual ex-post compensating variations for the
price changes under study. Conversely, every full-consideration model in
this preference class is a special case of the stable limited-consideration
model. Consequently, the two model classes generate exactly the same
structural demand functions on the maintained domain.
\end{proposition}

\noindent \textbf{Proof.} Let $h:\mathbb{R}\rightarrow (0,1)$ be continuous
and strictly increasing, for example $h(v)=(1+\exp (-v))^{-1}$. Let $g:%
\mathbb{R}\rightarrow (-2,-1)$ be continuous and strictly increasing. For
each $\widetilde{\eta }$, define for each $j\in \mathcal{J}$,%
\begin{equation}
\widetilde{U}_{j}(m;\widetilde{\eta })=\left\{ 
\begin{array}{ll}
h(U_{j}(m;\eta )), & j\in S, \\ 
g(m), & j\notin S.%
\end{array}%
\right.  \label{eq:equivalent-utilities}
\end{equation}%
Every function in (\ref{eq:equivalent-utilities}) is finite, continuous,
strictly increasing in residual income, and measurable in the augmented
tastes. Utilities of considered alternatives lie in $(0,1)$, whereas those
of excluded alternatives lie in $(-2,-1)$. Since $S$ is nonempty, no
excluded alternative can maximize utility under full consideration.
Moreover, the common strictly increasing transformation $h$ preserves all
rankings within $S$. Thus, pointwise in tastes and configurations, 
\begin{equation}
\mathop{\rm arg\,max}_{j\in \mathcal{J}}\widetilde{U}_{j}(y-p_{j};\widetilde{%
\eta })=\mathop{\rm arg\,max}_{j\in S}U_{j}(y-p_{j};\eta ).
\label{eq:equivalent-choices}
\end{equation}%
Thus, maximization of the constructed utilities $\widetilde{U}$ over the
full choice set $\mathcal{J}$ reproduces the same structural demand as
maximization of the original utilities $U$ over the consideration set $S$.

The full-consideration indirect utility satisfies 
\begin{equation}
\widetilde{V}(p,y;\widetilde{\eta }):=\max_{j\in \mathcal{J}}\widetilde{U}%
_{j}(y-p_{j};\widetilde{\eta })=h(V_{S}(p,y;\eta )).
\label{eq:equivalent-indirect}
\end{equation}%
Since $h$ is one-to-one, for every transfer $c$, 
\begin{equation*}
\widetilde{V}(p^{\prime },y+c;\widetilde{\eta })=\widetilde{V}(p,y;%
\widetilde{\eta })\quad \Longleftrightarrow \quad V_{S}(p^{\prime },y+c;\eta
)=V_{S}(p,y;\eta ).
\end{equation*}%
Existence and uniqueness from Proposition \ref{prop:support} imply identical
individual compensation, and hence identical population distributions. The
converse follows by setting $S=\mathcal{J}$ almost surely. \hfill $\square $

The proposition implies equality of the classes of demand functions
generated by stable limited and full consideration. It does not imply that
any demand system is admissible. For example, holding income and other
prices fixed, raising the price of alternative $j$ cannot increase its
choice probability. Empirical restrictions on demand therefore remain, even
though demand alone cannot distinguish exclusion from sufficiently
unfavorable preferences in this class.

The equivalence explains why the full-consideration welfare formula survives
stable limited consideration: both choices and ex-post compensation admit a
full-consideration representation.

\subsection{A boundary to equivalence: sufficiently large subsidies}

Proposition \ref{prop:equivalence} allows an alternative to remain less
attractive than its competitors at every residual income. Allowing residual
income to become arbitrarily large is not itself sufficient to rule this
out: strictly increasing utilities can remain bounded. An additional
preference restriction ruling out this possibility can distinguish limited
from full consideration when sufficiently large subsidies are observable.

\begin{proposition}[Limiting demand and the probability of consideration]
\label{prop:limiting-consideration} Take alternative 1 to be an inside good,
distinct from the outside option. Fix income $y$ and all other prices $%
p_{-1} $, and write $p(t)$ for the price vector with $p_1(t)=t$ and these
other prices unchanged. Suppose Assumptions \ref{ass:preferences}--\ref%
{ass:noties} hold along this path, with each consumer's set $S$ and the
joint population distribution of $(\eta,S)$ fixed as $t\to-\infty$. Suppose
also that, for almost every $\eta$, alternative 1 eventually dominates every
other alternative: there exists a finite $M(\eta)$ such that 
\begin{equation}
U_1(m;\eta)>\max_{j\neq1}U_j(y-p_j;\eta) \quad\hbox{for all }m>M(\eta).
\label{eq:eventual-dominance}
\end{equation}
Then 
\begin{equation}
\lim_{t\to-\infty}q_1(p(t),y)=\Pr(1\in S).  \label{eq:limiting-consideration}
\end{equation}
In particular, the limit is one under full consideration, and is strictly
less than one if $\Pr(1\notin S)>0$. A sufficient condition for (\ref%
{eq:eventual-dominance}) is $\lim_{m\to\infty}U_1(m;\eta)=+\infty$ almost
surely.
\end{proposition}

\noindent \textbf{Proof.} If $1\notin S$, alternative 1 is never chosen. If $%
1\in S$, then $y-t\rightarrow \infty $ while the residual incomes of all
competitors remain fixed. By (\ref{eq:eventual-dominance}), alternative 1 is
eventually the unique maximizer over $S$. Hence, almost surely, 
\begin{equation*}
\mathbf{1}\{d(p(t),y;\eta ,S)=1\}\longrightarrow \mathbf{1}\{1\in S\}.
\end{equation*}%
The indicators are bounded by one, so dominated convergence proves (\ref%
{eq:limiting-consideration}). Finally, the maximum of the finitely many
competitors' utilities at their fixed residual incomes is finite. Unbounded
utility for alternative 1 therefore implies eventual dominance. \hfill $%
\square $

\begin{corollary}[When limiting demand rules out full consideration]
\label{cor:reject-full} Suppose a structural demand system $q$ admits a
stable limited-consideration rationalization satisfying the conditions of
Proposition \ref{prop:limiting-consideration}. If 
\begin{equation*}
L:=\lim_{t\rightarrow -\infty }q_{1}(p(t),y)<1,
\end{equation*}%
then every rationalization in this class satisfies 
\begin{equation}
\Pr (1\notin S)=1-L>0,\qquad \Pr (S\neq \mathcal{J})\geq 1-L.
\label{eq:exclusion-share}
\end{equation}%
No full-consideration model satisfying the same eventual-dominance
restriction and fixed-population condition can rationalize $q$.
\end{corollary}

\noindent \textbf{Proof.} In every admissible limited-consideration
rationalization, Proposition \ref{prop:limiting-consideration} gives $\Pr
(1\in S)=L$. Exclusion of alternative 1 implies $S\neq \mathcal{J}$, which
yields (\ref{eq:exclusion-share}). If an admissible full-consideration model
generated the same demand, eventual dominance would make alternative 1
uniquely optimal for almost every consumer at all sufficiently negative own
prices. Dominated convergence would then imply $\lim_{t\rightarrow -\infty
}q_{1}(p(t),y)=1$, contradicting $L<1$. \hfill $\square $

The above argument does not require eliminating income effects. For example,
take two alternatives with $p_{0}=0$, $p_{1}=t$, $U_{0}(m;\eta )=m$ and $%
U_{1}(m;\eta )=2m+\eta $, where $\eta $ has the standard logistic
distribution. Independently of $\eta $, let $S=\{0,1\}$ with probability $%
L\in (0,1)$ and $S=\{0\}$ otherwise. Then 
\begin{equation*}
q_{1}(p(t),y)=\frac{L}{1+\exp (2t-y)},\qquad q_{0}=1-q_{1}.
\end{equation*}%
These demands are rationalized by stable limited consideration, exhibit
income effects, and satisfy $\lim_{t\rightarrow -\infty }q_{1}=L<1$. Both
utility functions are unbounded above. Thus no full-consideration model
imposing eventual dominance for alternative 1 can reproduce these demands
along the entire price path. Proposition \ref{prop:equivalence} still
supplies a full-consideration representation in the broader class that
permits permanent inferiority of an alternative.

\section{Extension and Illustration}

\subsection{Welfare bounds under contracting consideration}

\label{sec:monotone-bounds}

The preceding results hold under stable consideration. We now illustrate how
to accommodate one possible deviation from this assumption, viz. by allowing
component-wise price increase to weakly contract each consumers'
consideration sets. This will happen if, for example, online consumers
filter results by price and only consider options below a personal price
threshold.

Let $\xi =(\eta ,\nu )$ collect preference and consideration heterogeneity,
with a fixed population distribution. There are finitely many alternatives $%
j\in \mathcal{J}$. As before, each $U_{j}(m;\eta )$ is finite, continuous,
and strictly increasing in residual income $m$. Consideration is represented
by a nonempty set $S(p;\xi )$, which may depend on prices and be arbitrarily
correlated with preferences. Let $p^{0}\leq p^{1}$ componentwise and write 
\begin{equation}
\Delta _{j}=p_{j}^{1}-p_{j}^{0}\geq 0,\qquad j\in \mathcal{J}.
\label{eq:mb-increments}
\end{equation}%
The outside option $o$, priced at zero in both regimes, belongs to every
consideration set.

\begin{assumption}[Individual-level monotone contraction]
\label{ass:mb-contraction} For almost every $\xi$, and all $p,\widetilde p\in%
[p^0,p^1]$, 
\begin{equation}
p\leq\widetilde p\quad\Longrightarrow\quad S(\widetilde p;\xi)\subseteq
S(p;\xi).  \label{eq:mb-nesting}
\end{equation}
\end{assumption}

Write $S^r=S(p^r;\xi)$ for $r=0,1$ and $V_s(p,z;\eta)=\max_{j\in
s}U_j(z-p_j;\eta)$. We evaluate the compensation required to restore
baseline utility when the consumer faces the \emph{final} consideration set: 
\begin{equation}
V_{S^1}(p^1,y+C;\eta)=V_{S^0}(p^0,y;\eta).  \label{eq:mb-cv}
\end{equation}
The sets $S^0$ and $S^1$ are held fixed when varying the compensating
transfer. Assume unique maximizing alternatives almost surely at baseline
and at each constructed budget, and assume, almost surely, 
\begin{equation}
\lim_{c\to-\infty}V_{S^1}(p^1,y+c;\eta) <V_{S^0}(p^0,y;\eta)
<\lim_{c\to\infty}V_{S^1}(p^1,y+c;\eta).  \label{eq:mb-range}
\end{equation}
These conditions ensure a unique finite $C$. Since $S^1\subseteq S^0$ and
prices weakly rise, $C\geq0$.

\subsection{Demand and the attrition bound}

For $a\geq0$, define the constructed price--income configuration 
\begin{equation}
J_a=\{j:\Delta_j\leq a\},\qquad r_j(a)=p_j^0+\min\{\Delta_j,a\},\qquad
z(a)=y+a.  \label{eq:mb-budget}
\end{equation}
By monotone contraction, 
\begin{equation}
S^1\subseteq S^a\subseteq S^0, \qquad S^a:=S(r(a);\xi).
\label{eq:mb-intermediate-set}
\end{equation}
Let structural demand under price-responsive consideration be 
\begin{equation}
q_j^{\mathrm{obs}}(p,z)= \Pr\left\{j=\arg\max_{k\in
S(p;\xi)}U_k(z-p_k;\eta)\right\}.  \label{eq:mb-demand}
\end{equation}
Define the demand functional and attrition probability by 
\begin{equation}
B(a)=\sum_{j\in J_a}q_j^{\mathrm{obs}}(r(a),y+a), \qquad \ell=\Pr(S^0\neq
S^1).  \label{eq:mb-demand-bound}
\end{equation}
Assume the demand entering $B(a)$ is identified and an upper bound $%
\bar\ell\geq\ell$ is available, either from auxiliary data or as a
sensitivity parameter.

\begin{proposition}[Bounds on the compensation distribution]
\label{prop:mb-bounds} Under the preceding assumptions, $F_C(a)=0$ for $a<0$%
, and, for every $a\geq0$, 
\begin{equation}
\max\{0,B(a)-\bar\ell\}\leq F_C(a)\leq B(a).  \label{eq:mb-cdf-bounds}
\end{equation}
Because the outside option is always considered, the lower bound strengthens
to 
\begin{equation}
\max\{q_o^{\mathrm{obs}}(p^0,y),B(a)-\bar\ell\} \leq F_C(a)\leq B(a),\qquad
a\geq0.  \label{eq:mb-outside-bounds}
\end{equation}
\end{proposition}

\noindent\textbf{Proof.} Fix $a\geq0$ and set $b_j=U_j(y-p_j^0;\eta)$ and $%
n_j=U_j(y+a-p_j^1;\eta)$. The constructed utility satisfies 
\begin{equation}
h_j:=U_j(y+a-r_j(a);\eta) =%
\begin{cases}
n_j, & j\in J_a, \\ 
b_j, & j\notin J_a,%
\end{cases}
=\max\{b_j,n_j\}.  \label{eq:mb-max-utility}
\end{equation}
Let $j^*$ uniquely maximize $h_j$ over $S^a$ and define $E_a=\{j^*\in J_a\}$%
, $A_a=\{C\leq a\}$, and $T=\{S^0\neq S^1\}$. Then $\Pr(E_a)=B(a)$ and $%
\Pr(T)=\ell$.

We first show $A_a\subseteq E_a$ almost surely. If $j^*\notin J_a$, then $%
h_{j^*}=b_{j^*}>n_{j^*}$. For every $k\in S^1\subseteq S^a$ other than $j^*$%
, uniqueness gives $n_k\leq h_k<h_{j^*}$; the same strict inequality for $%
n_{j^*}$ applies if $j^*\in S^1$. Since the choice set is finite, 
\begin{equation*}
V_{S^1}(p^1,y+a;\eta)<h_{j^*}=b_{j^*} \leq V_{S^0}(p^0,y;\eta), 
\end{equation*}
so $C>a$. Next, on $T^c$, nesting implies $S^0=S^a=S^1$. If $E_a$ also
occurs, then 
\begin{equation*}
V_{S^1}(p^1,y+a;\eta)\geq n_{j^*}=h_{j^*} =\max_{j\in S^0}h_j\geq\max_{j\in
S^0}b_j, 
\end{equation*}
and therefore $C\leq a$. Thus $E_a\setminus T\subseteq A_a\subseteq E_a$,
which implies $B(a)-\ell\leq F_C(a)\leq B(a)$.

Finally, a consumer initially choosing the outside option has baseline
utility $U_o(y;\eta)$. This utility remains attainable at final prices
without compensation, while contraction and weakly higher prices cannot
increase indirect utility. Hence such a consumer has $C=0$, giving $%
F_C(a)\geq q_o^{\mathrm{obs}}(p^0,y)$ for $a\geq0$. Using $\ell\leq\bar\ell$
proves the result. \hfill$\square$

If every inside-option price rises strictly, the no-ties condition also
gives 
\begin{equation}
F_{C}(0)=q_{o}^{\mathrm{obs}}(p^{0},y).  \label{eq:mb-zero-compensation}
\end{equation}%
When $\bar{\ell}=0$, Proposition~\ref{prop:mb-bounds} recovers the
stable-consideration equality. More generally, $B(a)$ need not be
nondecreasing in $a$, because the constructed prices may induce changes in
consideration. Applying the monotonicity of the true CDF yields the
following refinement: 
\begin{equation}
\max \left\{ q_{o}^{\mathrm{obs}}(p^{0},y),\sup_{0\leq t\leq a}B(t)-\bar{\ell%
}\right\} \leq F_{C}(a)\leq \inf_{t\geq a}B(t),\qquad a\geq 0.
\label{eq:monotone-envelope}
\end{equation}

\begin{remark}
We do not pursue a general sharpness characterization of these bounds. Our
assumption that consideration responds to prices only is restrictive, as it
may also depend on information, search costs, and other factors.
Establishing sharpness under this particular specification may therefore
have limited economic value without a better understanding of the
determinants of consideration. Instead, we interpret the bounds as a
sensitivity analysis, quantifying how welfare conclusions under stable
consideration may change when a specified fraction of consumers lose
considered alternatives as prices rise.
\end{remark}

\begin{remark}
Under stable consideration, $C\leq D:=\max_j\Delta_j$. With contracting
consideration, this ceiling applies to consumers whose sets do not change, a
fraction of at least $1-\bar\ell$. Consumers losing a highly valued
alternative may require compensation exceeding $D$ by an arbitrarily large
amount. Accordingly, the bounds guarantee $F_C(D)\geq1-\bar\ell$, but do not
furnish a finite uniform upper bound on mean compensation.
\end{remark}

\subsection{Quantile implications}

Let $q_o^0=q_o^{\mathrm{obs}}(p^0,y)$ and define 
\begin{equation}
L(a)=\max\left\{q_o^0,\sup_{0\leq t\leq a}B(t)-\bar\ell\right\}, \qquad
U(a)=\inf_{t\geq a}B(t).  \label{eq:quantile-envelopes}
\end{equation}
Both functions are nondecreasing and satisfy $L\leq F_C\leq U$. For $0<\tau<1
$, write $Q_H(\tau)=\inf\{a\geq0:H(a)\geq\tau\}$, with $\inf\varnothing=+%
\infty$, and $Q_C=Q_{F_C}$.

\begin{corollary}[Quantile bounds]
\label{cor:contraction-quantiles} For every $0<\tau<1$, 
\begin{equation}
Q_U(\tau)\leq Q_C(\tau)\leq Q_L(\tau).  \label{eq:quantile-bounds}
\end{equation}
Moreover, $Q_C(\tau)\leq D$ for $\tau\leq1-\bar\ell$, and $Q_C(\tau)=0$ for $%
\tau\leq q_o^0$.
\end{corollary}

\noindent\textbf{Proof.} The first inequalities follow by inversion of $%
L\leq F_C\leq U$. Consumers with unchanged consideration require at most $D$%
, so $F_C(D)\geq1-\bar\ell$. Initial outside-option choosers require zero
compensation, so $F_C(0)\geq q_o^0$. \hfill$\square$

Thus, attrition of at most ten percent bounds quantiles through the
ninetieth percentile by the largest price increase, but does not bound the
upper tail or mean compensation uniformly.

\subsection{Calibrating attrition from marginal consideration rates}

The attrition bound can be informed by auxiliary information without
identifying preferences. Let 
\begin{equation}
\pi_j^r=\Pr(j\in S^r),\quad r=0,1, \qquad \delta_j=\pi_j^0-\pi_j^1.
\label{eq:mb-marginal-rates}
\end{equation}

\begin{proposition}[Attrition from marginal consideration rates]
\label{prop:mb-attrition} If $S^1\subseteq S^0$ almost surely and the
outside option is always considered, then $\delta_j\geq0$ and 
\begin{equation}
\max_{j\neq o}\delta_j\leq\ell \leq\min\left\{1,\sum_{j\neq
o}\delta_j\right\}.  \label{eq:mb-attrition-marginals}
\end{equation}
Consequently, valid upper bounds $\bar\delta_j\geq\delta_j$ imply the valid
attrition bound $\bar\ell=\min\{1,\sum_{j\neq o}\bar\delta_j\}$.
\end{proposition}

\noindent\textbf{Proof.} Nesting gives $\delta_j=\Pr(j\in S^0\setminus S^1)$%
. The attrition event is the union of these alternative-specific loss
events. The union bound and its corresponding lower bound prove the result.
\hfill$\square$

Representative surveys of consideration before and after the price change
can inform $\bar\ell$ without matching individuals across regimes.

\subsection{Why small attrition can conceal large welfare losses}

The following example shows that welfare uncertainty can persist even with
full structural demand and a known attrition rate.

\begin{example}[Observationally equivalent attrition losses]
\label{ex:binary-attrition} There are two alternatives, with $U_0(m)=m$ and $%
U_1(m;v)=m+v$. The outside price is zero, and the inside price rises from
zero to $d>0$. A fraction $1-\ell$ always considers both options and has $%
v=v_s>d$. The remaining fraction $0<\ell<1$ has $v=M>0$ and considers the
inside option only when its price is zero. This consideration rule is income
invariant and contracts monotonically.

At every income, the inside option is chosen by everyone at price zero and
by exactly the stable fraction $1-\ell$ at every positive price in $(0,d]$.
Thus the entire structural demand system on this domain, and the attrition
rate, are independent of $M$. Nevertheless, compensation is $d$ for stable
consumers and $M$ for consumers who lose the inside option: 
\begin{equation}
F_C(a)=(1-\ell)\mathbf{1}\{d\leq a\} +\ell\mathbf{1}\{M\leq a\},\qquad 
\mathbb{E}C=(1-\ell)d+\ell M.  \label{eq:binary-attrition-cdf}
\end{equation}
For $M>d$, every quantile above $1-\ell$ equals $M$. Allowing $M$ to
increase therefore makes these quantiles and mean compensation arbitrarily
large without changing demand or attrition.

Here $B(0)=0$, $B(a)=\ell$ for $0<a<d$, and $B(a)=1$ for $a\geq d$. For each
fixed $a>0$, choosing $0<M\leq a$ or $M>a$ attains, respectively, the upper
or lower pointwise CDF bound when $\bar\ell=\ell$ (with the stable
contribution included when $a\geq d$).
\end{example}

This is pointwise attainability for one demand system, not a general
sharpness result: the observationally equivalent economies vary the
preference parameter $M$. The example does not establish attainability with
the preference distribution fixed.

\subsection{A worked illustration}

We end the paper with a concrete illustration of how to implement our
methods, taking withdrawal of US clean-vehicle credits as a hypothetical
example. The credits ended for vehicles acquired after 30 September 2025
(cf. Internal Revenue Service, 2026). The calculation below is an
illustration of how to estimate the withdrawal's direct effects (i.e.
ignoring pollution effects) on consumer welfare.

Consider eligible buyers choosing among three vehicles and an
always-considered outside option, such as retaining their existing vehicle.
Index the outside option by $0$, with $p_0=0$, an unsubsidized vehicle by $1$%
, and two subsidized vehicles by $2$ and $3$. Hold manufacturer prices,
vehicle attributes, and the population fixed, and suppose credit withdrawal
raises net purchase prices by 
\begin{equation*}
(\Delta_0,\Delta_1,\Delta_2,\Delta_3)=(0,0,3750,7500).
\end{equation*}
These are illustrative price increments; an empirical application would use
increments incorporating pass-through and any maintained supplier response.
Buyers may consider different, unobserved subsets of the vehicles.

Under stable consideration, Theorem \ref{thm:cdf} gives 
\begin{equation}
F_C(a)=\left\{%
\begin{array}{ll}
0, & a<0, \\[3pt] 
\displaystyle\sum_{j=0}^{1}q_j(r(a),y+a), & 0\leq a<3750, \\[5pt] 
\displaystyle\sum_{j=0}^{2}q_j(r(a),y+a), & 3750\leq a<7500, \\[5pt] 
1, & a\geq7500,%
\end{array}%
\right.  \label{eq:ev-policy-cdf}
\end{equation}
where 
\begin{equation*}
r(a)=(0,p_1,p_2+\min\{3750,a\},p_3+\min\{7500,a\}),\qquad a\geq0.
\end{equation*}

For example, the fraction compensated by \$5,000 is 
\begin{equation}
F_C(5000)=\sum_{j=0}^{2} q_j\big((0,p_1,p_2+3750,p_3+5000),y+5000\big).
\label{eq:ev-five-thousand}
\end{equation}
At this constructed budget, alternatives $0$, $1$, and $2$ deliver their
post-withdrawal residual incomes with \$5,000 compensation. Alternative $3$
retains its baseline residual income. Choosing one of the first three
therefore reveals that compensation is sufficient; choosing alternative $3$
reveals that it is insufficient. The classification holds for each buyer's
fixed consideration set, without the analyst observing that set.

Implementation requires structural demand at the constructed prices and
income for the same population of eligible buyers. These configurations
generally differ from both policy endpoints. The empirical design must
therefore provide variation covering the constructed path or justify
extrapolation, subject to the conditions in Section \ref%
{subsec:demand-identification}. The theorem identifies the welfare
functional without separately recovering buyers' vehicle shortlists.

If consideration instead contracts with prices as specified in Section \ref%
{sec:monotone-bounds}, the same constructed-demand expression provides
bounds rather than point identification. Let 
\begin{equation*}
B(a)=\sum_{j:\,\Delta _{j}\leq a}q_{j}^{\mathrm{obs}}(r(a),y+a),\qquad a\geq
0,
\end{equation*}%
where structural demand allows consideration to respond to prices. If at
most ten percent of consumers lose a considered alternative, Proposition \ref%
{prop:mb-bounds} implies 
\begin{equation*}
\max \{q_{0}^{\mathrm{obs}}(p^{0},y),B(a)-0.10\}\leq F_{C}(a)\leq B(a).
\end{equation*}%
Because $B$ need not be nondecreasing, let $L$ and $U$ denote the
monotone-envelope bounds defined in Section \ref{sec:monotone-bounds}.
Writing $Q_{H}(\tau )=\inf \{a\geq 0:H(a)\geq \tau \}$, the corresponding
quantile restrictions are 
\begin{equation*}
L(a)\leq F_{C}(a)\leq U(a),\qquad Q_{U}(\tau )\leq Q_{C}(\tau )\leq
Q_{L}(\tau ).
\end{equation*}%
In particular, since the largest price increase is \$7,500, at least 90
percent of consumers require compensation no greater than this amount: 
\begin{equation*}
Q_{C}(\tau )\leq \$7{,}500,\qquad 0<\tau \leq 0.90.
\end{equation*}%
Thus, a bound on consideration-set attrition restricts most of the welfare
distribution even when consideration responds to prices, although mean
compensation need not have a finite uniform upper bound. The calculation
illustrates a sensitivity analysis around the stable-consideration result.

\clearpage
\appendix
\numberwithin{equation}{section} \setcounter{theorem}{0} \renewcommand{%
\thetheorem}{\Alph{section}.\arabic{theorem}}

\section{Ex-ante and ex-post compensation}

\label{app:compensation}

Let $\eta $ denote preference heterogeneity, with distribution $G$, and let $%
S$ be a nonempty random consideration set taking values in $\mathcal{S}$,
with conditional probabilities $\pi (s\mid \eta )=\Pr (S=s\mid \eta )$;
consideration and preferences need not be independent. For a price vector $p$
and income $y$, write $V_{s}(p,y,\eta )=\max_{j\in s}U_{j}(y-p_{j},\eta )$.
Consider a price change from $p^{0}$ to $p^{1}$, holding the conditional
consideration mechanism fixed. Assume that utilities are continuous and
strictly increasing in the numeraire, that the expectations below are
finite, and that the compensating transfers exist. Ex-post compensation is
defined after the consideration set is realized, holding that same set fixed
in the before--after comparison: $C^{\mathrm{post}}(s,\eta )$ is the unique
transfer satisfying 
\begin{equation}
V_{s}(p^{1},y+C^{\mathrm{post}}(s,\eta ),\eta )=V_{s}(p^{0},y,\eta ).
\end{equation}%
Its population distribution averages over the joint distribution of
consideration and preferences: 
\begin{equation}
F_{\mathrm{post}}(c)=\int \sum_{s\in \mathcal{S}}\pi (s\mid \eta )\mathbf{1}%
\{C^{\mathrm{post}}(s,\eta )\leq c\}\,dG(\eta ).
\end{equation}%
Equivalently, the indicator in this expression is $\mathbf{1}%
\{V_{s}(p^{1},y+c,\eta )\geq V_{s}(p^{0},y,\eta )\}$.

In contrast, ex-ante compensation corresponds to before consideration is
realized, conditional on the consumer's preference type, and cannot depend
on the subsequently realized set. Under expected-utility evaluation of the
consideration lottery, define $W(p,y,\eta )=\sum_{s\in \mathcal{S}}\pi
(s\mid \eta )V_{s}(p,y,\eta )$. The unique ex-ante transfer $C^{\mathrm{ante}%
}(\eta )$ satisfies 
\begin{equation}
W(p^{1},y+C^{\mathrm{ante}}(\eta ),\eta )=W(p^{0},y,\eta ),
\end{equation}%
and its population distribution is 
\begin{eqnarray}
F_{\mathrm{ante}}(c) &=&\int \mathbf{1}\{C^{\mathrm{ante}}(\eta )\leq
c\}\,dG(\eta )  \notag \\
&=&\int \mathbf{1}\{W(p^{1},y+c,\eta )\geq W(p^{0},y,\eta )\}\,dG(\eta ).
\end{eqnarray}%
Thus ex-post compensation varies with both the realized consideration set
and preferences, whereas ex-ante compensation varies only with preferences
after averaging utility over conditional consideration outcomes. In general, 
$C^{\mathrm{ante}}(\eta )\neq \sum_{s\in \mathcal{S}}\pi (s\mid \eta )C^{%
\mathrm{post}}(s,\eta )$: compensation for expected utility is distinct from
expected ex-post compensation.

\textbf{Remark}: \textit{Unlike the ex-post measure, the ex-ante measure
requires a cardinal utility representation of preferences over consideration
lotteries and is not invariant to arbitrary increasing transformations of
the utility functions. The main text therefore focuses on ex-post
compensation.}

\clearpage

\section{Adding and removing options}

\label{sec:menus}

The fixed-set theorem evaluates changes in prices for fixed but unknown
consideration sets. This section instead considers a specified change in the
menu of options that can enter consideration. The objective is to show how
to compute welfare effects when consideration itself changes. The result
uses demand under the union of the initial and final menus. It covers
multiple additions and removals, including simultaneous changes in prices of
retained alternatives, but requires more than demand under an unchanged
baseline set.

Let $M_{0}$ and $M_{1}$ be known initial and final menus, and write 
\begin{equation*}
K=M_{0}\cup M_{1},\qquad H=M_{0}\cap M_{1},\qquad R=M_{0}\setminus
M_{1},\qquad N=M_{1}\setminus M_{0}.
\end{equation*}%
For each consumer, let $S\subseteq K$ be an unobserved potential
consideration set, with arbitrary dependence on preferences. The realized
sets in the two regimes are $S\cap M_{0}$ and $S\cap M_{1}$, both assumed
nonempty. Thus the intervention changes which options can be considered,
without changing the consumer's inclusion of retained options. The set $S$
need not be identified. Under the union menu $K$, the consumer must consider
precisely $S$, including at the constructed budgets below.

Let $p_{j}^{0}$ be defined for $j\in M_{0}$ and $p_{j}^{1}$ for $j\in M_{1}$%
. Denote the outside option by $o\in H$ with $p_{o}^{0}=p_{o}^{1}=0$.
Preferences are unchanged across regimes. Define 
\begin{equation*}
V_{t}(z;\eta ,S)=\max_{j\in S\cap M_{t}}U_{j}(z-p_{j}^{t};\eta ),\qquad
t=0,1,
\end{equation*}%
and define $C_{01}$ by $V_{1}(y+C_{01};\eta ,S)=V_{0}(y;\eta ,S)$. Unlike a
fixed-set price change, a menu change need not admit finite compensation
under monotonicity alone. We therefore require, almost surely, 
\begin{equation}
\lim_{c\rightarrow -\infty }V_{1}(y+c;\eta ,S)<V_{0}(y;\eta
,S)<\lim_{c\rightarrow \infty }V_{1}(y+c;\eta ,S).  \label{eq:menu-range}
\end{equation}%
For example, this holds if each $U_{j}(\cdot ;\eta )$ maps $\mathbb{R}$ onto 
$\mathbb{R}$. The condition ensures existence of finite compensation.

Write $q_j^K(r,z)$ for structural choice probabilities under the union menu,
over the same population distribution of $(\eta,S)$. For a threshold $a$,
define the residual incomes 
\begin{equation}
m_j^*(a)= \left\{%
\begin{array}{ll}
\max\{y-p_j^0,y+a-p_j^1\}, & j\in H, \\ 
y-p_j^0, & j\in R, \\ 
y+a-p_j^1, & j\in N,%
\end{array}%
\right.  \label{eq:menu-residuals}
\end{equation}
and put $z(a)=y+\max\{a,0\}$ and $r_j(a)=z(a)-m_j^*(a)$. Then $r_o(a)=0$.
Finally, let 
\begin{equation*}
A(a)=N\cup\{j\in H:p_j^1-p_j^0\leq a\}.
\end{equation*}

\begin{theorem}[Compensation for additions and removals]
\label{thm:menus} Suppose preferences satisfy Assumption \ref%
{ass:preferences}, the menu structure and range condition above hold, and
the joint distribution of $(\eta ,S)$ is fixed across regimes and
constructed budgets. Suppose the maximizer over $S$ at each constructed
union-menu budget is unique almost surely. Then $C_{01}$ exists uniquely
and, for every $a\in \mathbb{R}$, 
\begin{equation}
\Pr (C_{01}\leq a)=\sum_{j\in A(a)}q_{j}^{K}(r(a),z(a)).  \label{eq:menu-cdf}
\end{equation}%
Consequently, structural demand under the union menu at these budgets
identifies the entire compensation distribution without identifying $S$.
\end{theorem}

\noindent \textbf{Proof.} Continuity, strict monotonicity, and (\ref%
{eq:menu-range}) give existence and uniqueness. Fix $a$ and write $%
B=V_{0}(y;\eta ,S)$ and $D=V_{1}(y+a;\eta ,S)$. At the constructed budget,
the maximum utility over $S$ is $\max \{B,D\}$. Outside a null event, let $%
j^{\ast }$ be its unique maximizer. If $j^{\ast }\in A(a)$, its constructed
utility is a final-regime utility and equals $D$, so $D\geq B$. If $j^{\ast
}\in R$, its constructed utility is a baseline utility. All final-regime
alternatives are distinct from $j^{\ast }$ and have utility strictly below
its constructed utility, giving $D<B$. In the remaining case $j^{\ast }\in
H\setminus A(a)$, its final utility is strictly below its baseline utility
by monotonicity; every other final-regime alternative $j$ has utility no
greater than its own constructed utility, which is strictly below that of $%
j^{\ast }$ by uniqueness. Again $D<B$. Thus $\{D\geq B\}=\{j^{\ast }\in
A(a)\}$ almost surely. Strict monotonicity gives $\{C_{01}\leq a\}=\{D\geq
B\}$. Taking probabilities proves (\ref{eq:menu-cdf}). \hfill $\square $

\subsection{Pure removal and pure addition}

With unchanged prices for retained goods, the formula has particularly
simple forms. For removal, let $M_1=M_0\setminus R$, so $K=M_0$, and write $%
p $ for baseline prices. Then $C_{01}\geq0$ and, for $a\geq0$, 
\begin{equation}
\Pr(C_{01}\leq a) =1-\sum_{j\in R}q_j^{M_0}(r^R(a),y+a),\qquad r_j^R(a)=
\left\{%
\begin{array}{ll}
p_j+a, & j\in R, \\ 
p_j, & j\in M_1.%
\end{array}%
\right.  \label{eq:removal-cdf}
\end{equation}
At this budget, removed options retain their baseline residual incomes,
while surviving options receive the candidate compensation. Demand under the
initial menu therefore suffices. The formula allows each consumer to have
considered only some, or none, of the removed options.

For addition, let $M_1=M_0\cup N$, so $K=M_1$, and write $p$ for final
prices, which equal baseline prices on $M_0$. Then $C_{01}\leq0$ and, for $%
a<0$, 
\begin{equation}
\Pr(C_{01}\leq a) =\sum_{j\in N}q_j^{M_1}(r^N(a),y),\qquad r_j^N(a)= \left\{%
\begin{array}{ll}
p_j-a, & j\in N, \\ 
p_j, & j\in M_0.%
\end{array}%
\right.  \label{eq:addition-cdf}
\end{equation}
For $a\geq0$ the CDF equals one. Equivalently, for $b>0$, the fraction
willing to pay at least $b$ for access to the added options equals their
combined demand under the expanded menu when all their prices rise by $b$.
Post-addition demand is required: baseline demand alone contains no
information about the attractiveness of options that nobody could consider.

The one-sided support allows recovery of mean losses or gains by integrating
the corresponding tail probabilities, provided these means are finite.
Unlike the fixed-set price-change case, compensation need not have bounded
support here.

\subsection{Identification and relation to price limits}

For pure removal the union menu is the initial menu; for pure addition it is
the final menu. For a mixed intervention, demand when both removed and added
options are available is an additional object. It need not be identified
from the two endpoint demand systems, and this theorem does not claim
otherwise. More generally, arbitrary person-specific changes in
consideration need not be representable as $S\cap M_t$ for known menus with
stable consideration under their union.

The theorem can describe a platform withdrawing a known collection of
listings or introducing new listings, provided retained-option consideration
is unchanged and the union-menu demand requirement holds.

Bhattacharya (2018, Section 2.2, Corollary 1, pp.~579--580; proof
pp.~604--605) obtains the distributions for elimination of one alternative
from a limiting price increase, assuming that the option becomes inferior to
a remaining one as its price tends to infinity. The reverse comparison
yields retrospective addition benefits with the appropriate CV/EV
interpretation. Theorem \ref{thm:menus} instead applies the maximum-budget
argument directly to the union menu. It handles several additions and
removals without taking a price limit in the CDF formula, while making
compensation existence and union-menu demand explicit. When $M_0=M_1$, it
reduces to the fixed-set price-change calculation.

\clearpage

\end{document}